\documentclass[onecolumn,showpacs,10pt]{revtex4-1}%
\usepackage{amsmath,amssymb,graphicx}
\usepackage{amsmath}
\usepackage{amsfonts}
\usepackage{amssymb}
\usepackage[normalem]{ulem}
\usepackage{graphicx}%
\providecommand{\U}[1]{\protect\rule{.1in}{.1in}}
\begin{document}

\title{High-speed computational ghost imaging with compressed sensing based on a convolutional neural network}
\author{Hao Zhang}

\affiliation{School of Physics and Physical Engineering, Qufu Normal University, Qufu 273165, China}
\author{Deyang Duan}
\affiliation{School of Physics and Physical Engineering, Qufu Normal University, Qufu 273165, China\\
Shandong Provincial Key Laboratory of Laser Polarization and Information
Technology, Research Institute of Laser, Qufu Normal University, Qufu 273165, China}

\begin{quote}
\begin{abstract}
Computational ghost imaging (CGI) has recently been intensively studied as an indirect imaging technique. However, the speed of CGI cannot meet the requirements of practical applications. Here, we propose a novel CGI scheme for high-speed imaging. In our scenario, the conventional CGI data processing algorithm is optimized to a new compressed sensing (CS) algorithm based on a convolutional neural network (CNN). CS is used to process the data collected by a conventional CGI device. Then, the processed data are trained by a CNN to reconstruct the image. The experimental results show that our scheme can produce high-quality images with much less sampling than conventional CGI. Moreover, detailed comparisons between the images reconstructed using our approach and with conventional CS and deep learning (DL) show that our scheme outperforms the conventional approach and achieves a faster imaging speed.

\end{abstract}
\maketitle
\end{quote}

\section{Introduction}

Ghost imaging is an indirect imaging technique based on quantum properties (e.g., quantum entanglement or intensity correlation) of the light field[1-3]. Compared to conventional
optical imaging techniques, ghost imaging requires two light beams: a reference
light beam, which never illuminates the object and is directly measured by a
detector with a spatial resolution (e.g., a charge-coupled device) and an
object light beam, which, after illuminating the object, is measured by a
bucket detector with no spatial resolution. By correlating the photocurrents
from the two detectors, the ghost image is retrieved.
Previous works show that ghost imaging has potential applications in remote
sensing[4,5], light detection and ranging (lidar)[6,7], medical imaging[8-10], and super-resolution imaging[11,12].
However, conventional ghost imaging requires two optical paths,
which severely limits its application. Fortunately, Shapiro creatively
introduced the concept of computational ghost imaging (CGI) in 2008[13]. In the CGI setup, the idle
light is obtained by calculation, so the reference light path is omitted in the
experimental apparatus[14]. Compared with conventional
ghost imaging, CGI is more suitable for application
in remote sensing, radar, and other fields.

After more than 10 years, CGI theory and experiments have matured. However, CGI is still in
the laboratory stage. One of the critical problems is that the imaging speed
cannot meet practical applications. Generally, to produce a
clear image, conventional CGI, including conventional ghost imaging, takes approximately 5 minutes, which
obviously cannot meet the requirements of practical application, especially those of moving
target imaging. How to improve the speed of ghost imaging is one of the key
factors for realizing its application. Compressed sensing (CS)[15-18] and deep learning (DL)[19-22]
greatly improve the imaging speed, but there is still a gap compared with the speed of
classical optical imaging.

In this article, we propose a novel CGI scheme with
CS based on a conventional neural network (CS-CNN) to improve the imaging speed. The setup is based on a
conventional CGI experimental apparatus. First,
the data collected by the CGI device are compressed by the conventional CS algorithm; then, the
processed data is trained to reconstruct the ghost image. This scheme combines
the advantages of CS with a low sampling rate and a
CNN for fast image reconstruction. Theoretical and
experimental results show that this scheme is significantly faster than
conventional CS and a conventional DL algorithm with a CNN under the condition of obtaining the
same quality image.

\section{Theory}
We use a conventional CGI experimental device in our
work. The setup is shown in Fig. 1. In the setup, a quasi-monochromatic
laser illuminates an object $T(\rho )$, and the reflected light carrying the
object information is modulated by a spatial light
modulator. A bucket detector collects the modulated light $E_{di}(\rho ,t)$.
Correspondingly, the calculated light $E_{ci}(\rho ^{^{\prime }},t)$ can be
obtained by diffraction theory. The object image can be
reconstructed by correlating the signal output by the bucket detector and
calculated signal[23-25]; i.e.,
\begin{equation}
G(\rho ,\rho ^{^{\prime }})=\left\langle \left\vert E_{di}(\rho
,t)\right\vert ^{2}\left\vert E_{ci}(\rho ^{^{\prime }},t)\right\vert
^{2}\right\rangle -\left\langle \left\vert E_{di}(\rho ,t)\right\vert
^{2}\right\rangle \left\langle \left\vert E_{ci}(\rho ^{^{\prime
}},t)\right\vert ^{2}\right\rangle
\end{equation}%
where $\left\langle \cdot \right\rangle $ stands for an ensemble average.
The subscript $i=1,2,\cdot \cdot \cdot ,n$ denotes the $i$th
measurement, and $n$ denotes the total number of measurements. For simplicity,
the object function $T(\rho )$ is contained in $E_{di}(\rho ,t)$.

\begin{figure}[h!]
\centering\includegraphics[width=12cm]{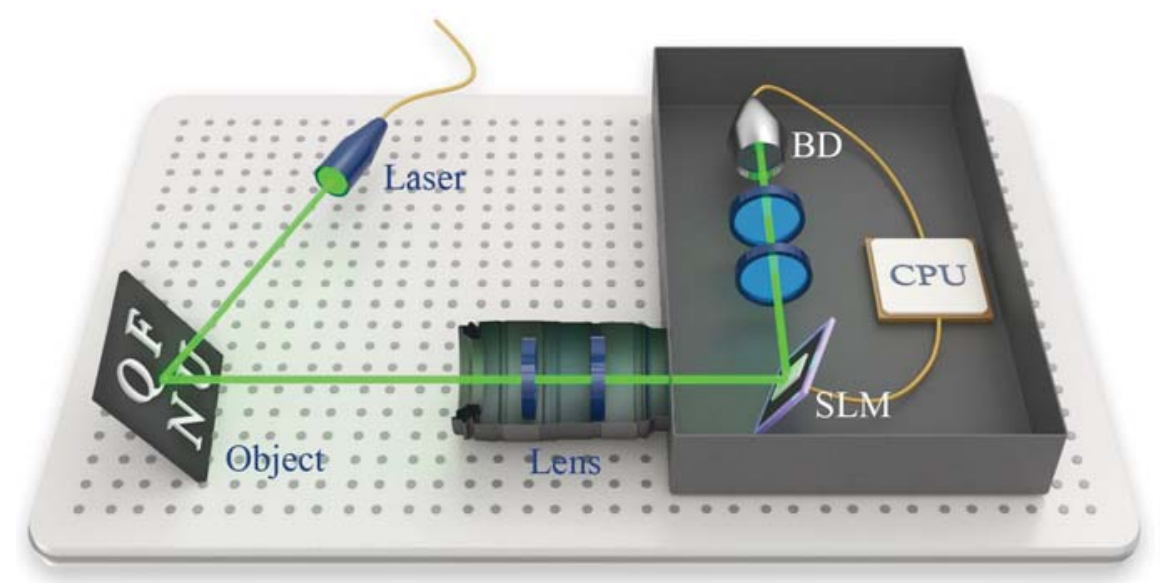}
\caption{Setup of the computational ghost imaging system with compressed sensing based
on a convolutional neural network. SLM: spatial light modulator, BD: bucket
detector.}
\end{figure}

The flow chart of the CS-CNN is shown in Fig. 2. In the
following, we briefly introduce the process of this algorithm. The algorithm
mainly consists of three parts: (i) a conventional CS program to
compress the data collected by the CGI device; (ii) a conventional CGI process program; and
(iii) a 10-layer CNN constructed for the training data.

\begin{figure}[h!]
\centering\includegraphics[width=13cm]{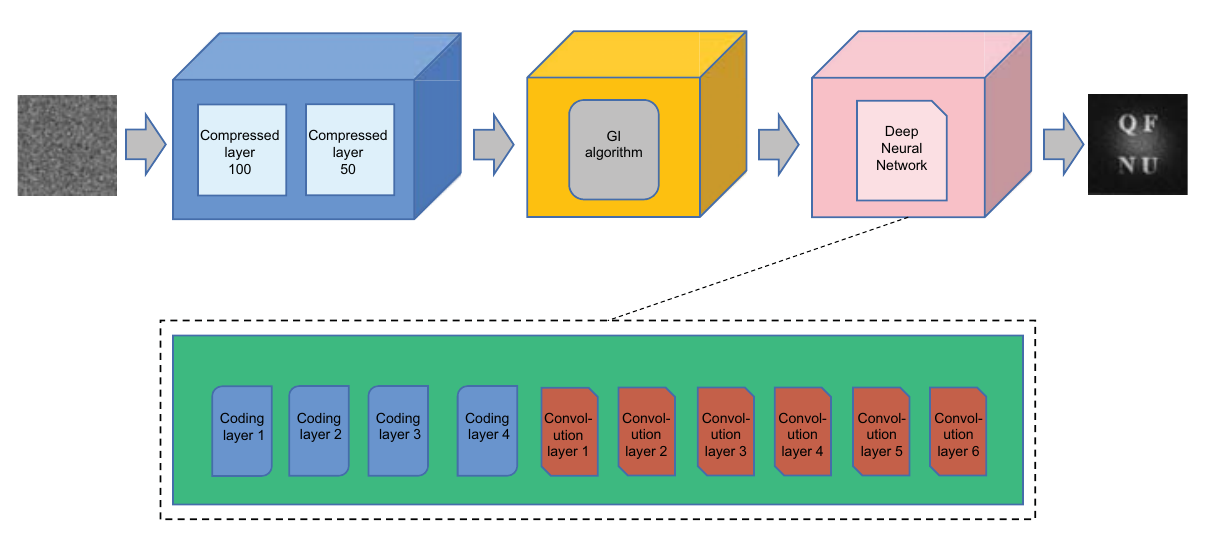}
\caption{Network structure of the proposed CS-CNN}
\end{figure}

In the conventional CGI device, a set of data ($n$) are measured by bucket
detector. Correspondingly, according to the diffraction theory of light,
the distribution of the idle light field in the object plane can be
obtained. Thus, we obtain $n$ 200$\times $200 data points. Each data point is divided
into 20$\times $20 blocks without overlapping. According to CS theory\cite{Katza15,Katkovnik16}, the random Gaussian matrix is used to process the data.
The rows of 20$\times $20 data blocks are arranged into a column vector to obtain
a 400-dimensional column vector. In this article, the measurement rate is $%
MR=0.25$, thus the size of the measurement matrix is 100$\times $400. Finally,
a 100-dimensional measurement vector is obtained. The above process can be
expressed as

\begin{equation}
y=\phi x
\end{equation}

\noindent where $\phi \in R^{M\times N}(M\ll N)$ is the measurement basis
matrix, $x\in R^{N}$ represents the vectorized image block, and $y\in R^{M}$ is
the measurement vector. $N/M$ represents the measurement rate. Following the
above steps, we can further compress the data to 50 dimensions.

A new set of data is obtained by processing the above data with a conventional
CGI program. Then, a 10-layer CNN
is constructed to train the data. Layers 1-4 of the network are stacked autoencoders, and layers 5-10 are convolution layers. The measurement matrix is
replaced by a stacked autoencoder, and the input layer is 20$%
\times $20 data blocks. All the rows are arranged into a 400$\times $%
1 column vector. If the number of neurons in the first layer is $C$, the measurement rate is $%
MR=C/400$. The first layer of the network is connected to the column
vector $x$ converted from the input image block, and the number of neurons $C
$ is set according to different measurement rates. The activation function
is a rectified linear unit (ReLU) function, which outputs the $C$-dimensional column vector $y$;
i.e.,

\[
y=T\left( W_{1}x+b_{1}\right)
\]

\noindent where $T$ represents the ReLU activation function and $W_{1}$ represents the weight parameter vector of neurons.
in the first layer, and $b_{1}$ represents the bias of neurons in the first
layer.

The second layer of the network is fully connected to the first layer,
which has 400 neurons. Take the output $y$ of the first layer as the input,
output $x$, and the activation function is the ReLU function. In the same
way, the third layer is fully connected to the second layer, with 100
neurons. The fourth layer is fully connected to the third layer, with 400
neurons. The initial reconstructed image block vector is rearranged into 20$%
\times $20 image blocks according to the original row and column to obtain
the preliminary reconstructed image block.

Finally, the CNN is used to reconstruct the image
block accurately. The output data of the fourth layer are taken as the input of
the fifth layer. In the fifth layer, 64 11$%
\times $11 convolution kernels are used to generate 64 10$\times $10 feature maps.
The sixth layer of the network is connected to the fifth layer (a convolution
layer), and 32 1$\times $1 convolution kernels are used to
generate 32 20$\times $20 characteristic graphs.
The seventh layer of the network is connected to the sixth layer (a convolution
layer), and a 7$\times $7 convolution kernel is used to
generate a 20$\times $20 feature map. The eighth layer of
the network is connected to the seventh layer (a convolution layer), and 64 11$\times $11
convolution cores are used to generate 64 20$\times $20feature maps.
The ninth layer of the network
is connected to the eighth layer (a convolution layer), and 32 1$\times $1 convolution
kernels are used to generate 32 20$\times $20 characteristic
graphs. The activation function of the above
process is a ReLU function. The tenth layer of the network is connected to the ninth
layer (a convolution layer). A 7$\times $7 convolution kernel
is used. The number of zeros in the tenth layer (a convolution layer) is 3,
and the output of the activation function is not used to generate the
reconstructed image block of size 20$\times $20.

In the deep learning framework Caffe, the 10-layer network is trained in an
unsupervised way, and the loss function is

\[
L(\{\mathrm{W}\})=\frac{1}{T}\sum_{1}^{T}\left\Vert F\left(
x_{i},\{W\}\right) -x_{i}\right\Vert ^{2}
\]%
\noindent where ${fan\underline{\hbox to 0.3cm{}}i_{in}}$ represents the
number of input units in the $i$th layer and ${fan\underline{\hbox to 0.3cm{}}%
i_{out}}$ represents the number of output neurons in the $i$th layer; the number
of input neurons in the first layer is 0, and the number of output neurons
in the fourth layer is 0. In the 5th to 10th layers of the network, the
initial weight distribution is subject to a Gaussian distribution with a
mean of 0 and a variance of 0.01. In layers 1-10 of the network, the
initial offset values are set to 0. After the deep neural network, the
reconstructed image blocks are obtained, and then the image blocks are
rearranged according to the original row,
and the row values are rearranged according to the index.
\section{Experiments and results}
\begin{figure}[h!]
\centering\includegraphics[width=12cm]{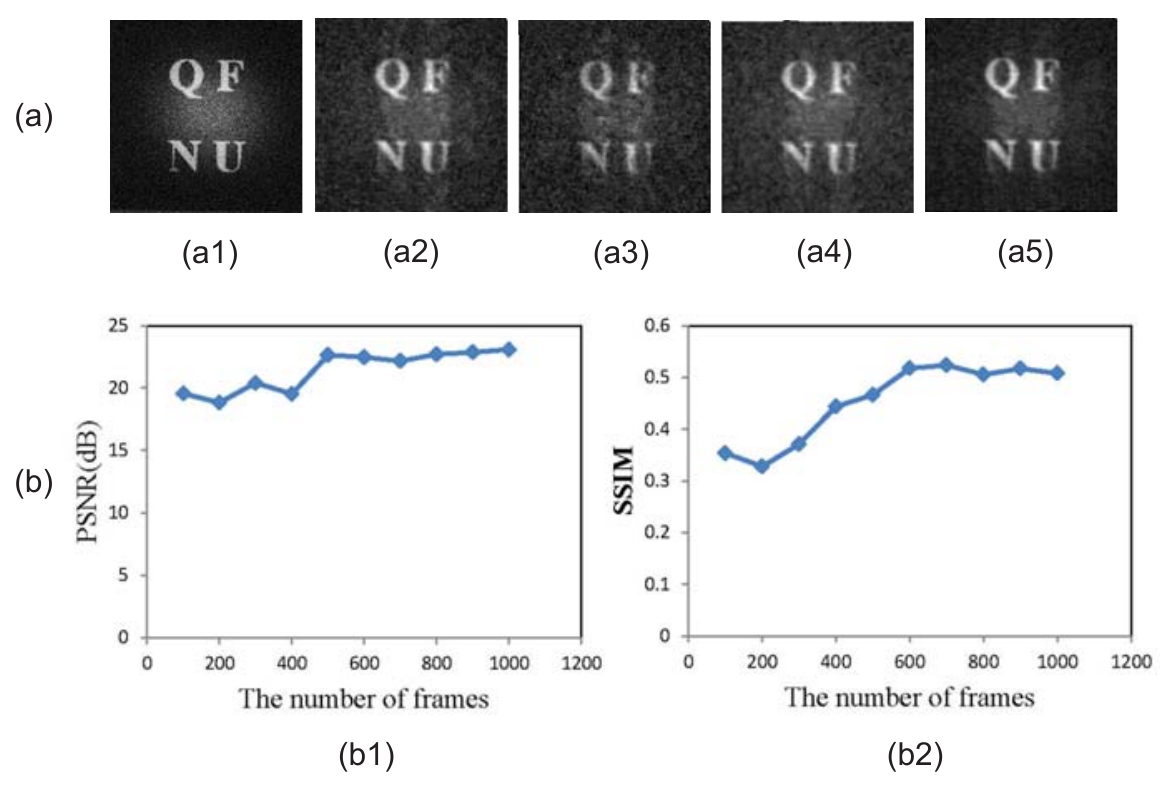}
\caption{The ghost image reconstructed by computational ghost imaging with compressed sensing based on a convolutional neural network. (a1) Classical image. The number of frames in the reconstructed ghost images are (a2) 100, (a3) 200, (a4)
300, (a5) 400. (b) The
PSNR and SSIM curves of reconstructed image with different frames number.}
\end{figure}
The experimental setup is schematically shown in
Fig. 1. A standard monochromatic laser
(30 mW, Changchun New Industries Optoelectronics Technology Co., Ltd. MGL-III-532)
with wavelength $\lambda
=532$ $nm$ illuminates an object
(Qufu Normal University, QFNU).
The light reflected by the object
focus on a two-dimensional amplitude-only ferroelectric liquid crystal
spatial light modulator (Meadowlark Optics A512-450-850) with 512$\times $%
512 addressable 15 $\mu m\times $15 $\mu m$ pixels through the lens. A bucket
detector collects the modulated light. Correspondingly, the reference signal is obtained by MATLAB software. The ghost image is reconstructed by the CS-CNN. In this experiment, the sampling rate is $MR=0.25$ and the number of training sets is 1000.

Fig. 3 shows a set of experimental results. Fig. 3(a1) is the object. Figs. 3(a2 - a5) represent reconstructed ghost images with different numbers of frames. The results show that the image quality is significantly improved by increasing the number of frames. High-quality ghost images comparable to classical optical imaging can be produced with little data. To quantitatively
analyze the quality of the reconstructed image at different frames, peak signal to noise
ratio (PSNR) and structural similarity index (SSIM) are used as our evaluation index. As can be
seen from Fig. 3(b), despite the number of samples is very small, the reconstructions are still in
reasonable quality.

\begin{figure}[h!]
\centering\includegraphics[width=12cm]{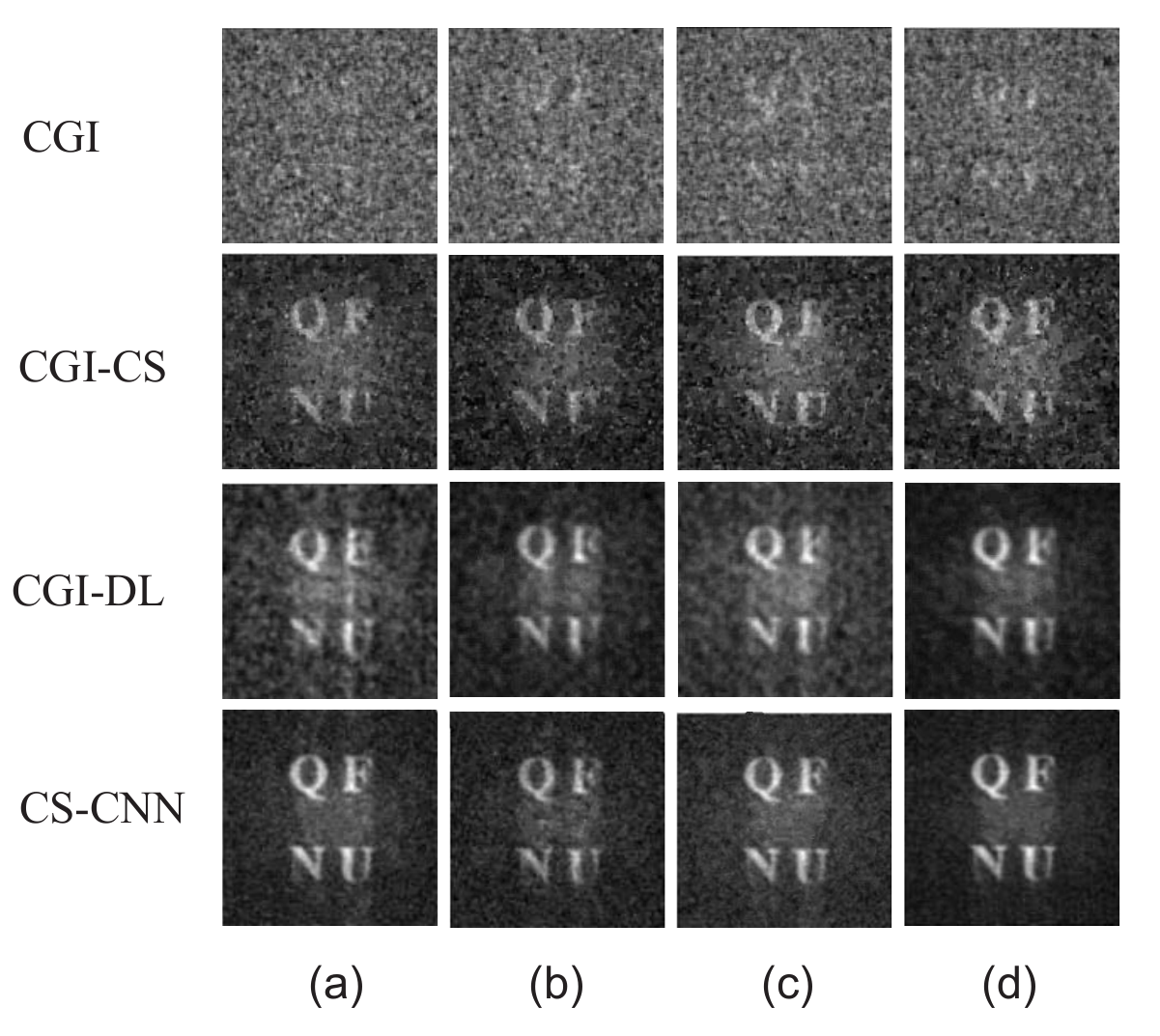}
\caption{Detailed comparisons between the ghost image reconstructed using the computational ghost imaging (CGI) algorithm, conventional compressed
sensing (CS) algorithm, deep learning (DL) algorithm and compressed
sensing algorithm based on a convolutional neural network (CS-CNN). The number of frames is (a) 100, (b) 200, (c) 200, and (d) 400.
}
\end{figure}
\begin{figure}[h!]
\centering\includegraphics[width=13cm]{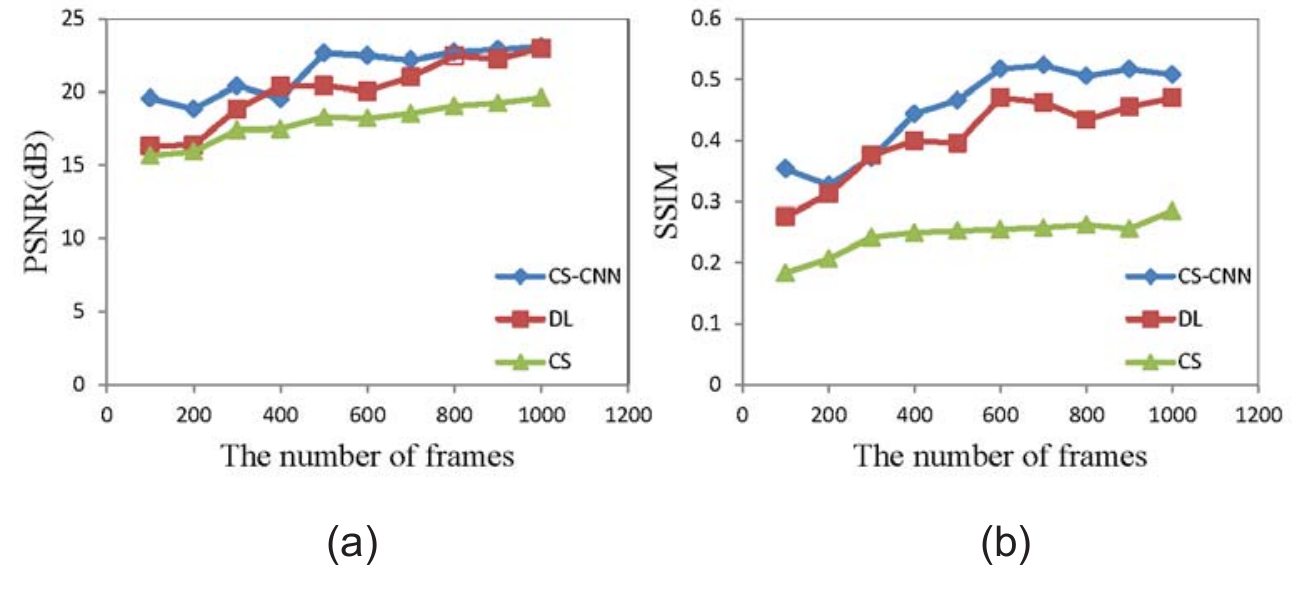}
\caption{The PSNR and SSIM curves of reconstructed images of CS, DL and CS-CNN with different frames number, respectively.
}
\end{figure}

We compare the conventional, CS, DL, and CS-CNN CGI algorithms based on the same experimental data. The conventional CS algorithm and CS-CNN algorithm have the same sampling rate; i.e., $MR=0.25$. The DL algorithm and CS-CNN algorithm set the same dataset. i.e., 1000. Fig. 4 shows that the frame number of the CS-CNN algorithm is obviously less than that of other algorithms to produce ghost images with the same quality. The quantitative results (Fig.5) show that the PSNR of CGI with CS-CNN is average 7\% higher than that of CGI with DL under the same reconstructed frame number, SSIM increased by 12\% on average[26].
Consequently, under the condition of producing the same image quality, CS-CNN has the faster imaging speed.
\section{Conclusions}
We have proposed a novel method for high-speed CGI. This method combines the advantages of the CS algorithm and CNN algorithm. We analyzed the performance of the conventional CGI, CS, and DL algorithms under the same conditions and observed that our CS-CNN scheme outperforms the other methods, especially when the sampling number is very small. To our knowledge, CS based on a CNN is the fastest CGI method to date. This method significantly reduces the data acquisition time in CGI, providing a promising solution to these challenges that prohibits the use of CGI in practical applications.
\section{Funding}
This project was supported by the National Natural Science Foundation (China)
under grant nos. 11704221, 11574178 and 61675115 and the Taishan Scholar
Project of Shandong Province (China) under grant no. tsqn201812059.
\section{Disclosures}
The authors declare that there are no conflicts of interest related to this article.

\bibliographystyle{unsrt}



\begin{thebibliography}{99}
\bibitem {revtex-au} T. B. Pittman, Y. H. Shih, D. V. Strekalov, and A. V. Sergienko.
 Optical imaging by means of two-photon quantum entanglement. Phys. Rev. A 52, R3429 (1995).

\bibitem {revtex-au} J. Cheng and S.-S. Han. Incoherent coincidence imaging and its applicability in X-ray diffractiont.
Phys. Rev. Lett. 92, 093903 (2004).

\bibitem {revtex-au} X. H. Chen, Q. Liu, K. H. Luo, and L. A. Wu.
Lensless ghost imaging with true thermal light.
Opt. Lett. 34, 695-697 (2009).

\bibitem {revtex-au}
 B. I. Erkmen.
Computational ghost imaging for remote sensing.
J. Opt. Soc. A 29, 782-789 (2012).

\bibitem {revtex-au}
 D. Y. Duan, Z. X. Man, and Y. J. Xia.
Nondegenerate wavelength computational ghost imaging with thermal light.
Opt. Express 27, 25187-25195 (2019).

\bibitem {revtex-au}
  C. Q. Zhao, W. L. Gong, M. L. Chen, E. R. Li, H. Wang, W. D. Xu, and S. S. Han.
Ghost imaging lidar via sparsity constraints.
Appl. Phys. Lett. 101, 141123 (2012).

\bibitem {revtex-au}
  W. L. Gong, C. Q. Zhao, H. Yu, M. L. Chen, W. D. Xu, and S. S. Han.
Three-dimensional ghost imaging lidar via sparsity constraint.
Sci. Rep. 6, 26133 (2016).

\bibitem {revtex-au}
  D. Pelliccia, A. Rack, M. Scheel, V. Cantelli, and D. M. Paganin.
Experimental X-Ray Ghost Imaging.
Phys. Rev. Lett. 117, 113902 (2016).

\bibitem {revtex-au}
  H. Yu, R. Lu, S. Han, H. Xie. G. Du, T. Xiao, and D. Zhu.
Fourier-Transform Ghost Imaging with Hard X Rays.
Phys. Rev. Lett. 117, 113901 (2016).

\bibitem {revtex-au}
   A. Zhang, Y. He, L. Wu, L. Chen and B. Wang.
Tabletop x-ray ghost imaging with ultra-low radiation.
Optica 5, 374-377 (2018).

\bibitem {revtex-au}
   W. Li, Z. Tong, K. Xiao, Z. Liu, Q. Gao, J. Sun, S. Liu, S. Han, and Z. Wang.
Single-frame wide-field nanoscopy based on ghost imaging via sparsity constraints.
 Optica 6, 1515-1523 (2019).

\bibitem {revtex-au}
   W. Gong and S. Han.
High-resolution far-field ghost imaging via sparsity constraint.
Sci. Rep. 5, 9280 (2015).

\bibitem {revtex-au}
   J. H. Shapiro.
Computational ghost imaging.
Phys. Rev. A 78, 061802(R) (2008).

\bibitem {revtex-au}
  Y. Bromberg, O. Katz, and Y. Silberberg.
Ghost imaging with a single detector.
Phys. Rev. A 79, 053840 (2009).

\bibitem {revtex-au}
  O. Katza, Y. Bromberg, and Y. Silberberg.
 Compressive ghost imaging.
 Appl. Phys. Lett. 95, 131110 (2009).

\bibitem {revtex-au}
  V. Katkovnik and J. Astola.
Compressive sensing computational ghost imaging.
 J. Opt. Soc. Am. A 29, 1556-1567 (2012).

\bibitem {revtex-au}
 W. K. Yu, M. F. Li, X. R. Yao, X. F. Liu, L. A. Wu, and G. J. Zhai .
 Compressive sensing computational ghost imaging.
Opt. Express 22, 7133-7144 (2014).

\bibitem {revtex-au}
  Z. Chen, J. Shi, and G. Zeng .
Object authentication based on compressive ghost imaging.
Appl. Opt. 55, 8644-8650 (2016).

\bibitem {revtex-au}
  M. Lyu, W. Wang, H. Wang, W. Wang, G. Li, N. Chen, and G. Situ.
Deep-learning-based ghost imaging .
Sci. Rep. 7, 17865 (2017).

\bibitem {revtex-au}
  Y. He, G. Wang, G. Dong, S. Zhu, H. Chen, A. Zhang, and Z. Xu.
Ghost imaging based on deep learning .
Sci. Rep. 8, 6469 (2018).

\bibitem {revtex-au}
  T. Shimobaba, Y. Endo, T. Nishitsuji, T. Takahashi, Y. Nagahama, T. Hasegawa, M. Sano, R. Hirayama, T. Kakue, A. Shiraki, and T. Ito .
Computational ghost imaging using deep learning .
Opt. Commun. 413, 147-151 (2018).

\bibitem {revtex-au}
 G. Barbastathis, A. Ozcan, and G. Situ .
 On the use of deep learning for computational imaging .
Optica 6, 921-943 (2019).

\bibitem {revtex-au}
 X. L. Yin, Y. J. Xia, and D. Y. Duan.
On the use of deep learning for computational imaging .
Opt. Express 26, 18944-18949 (2018).

\bibitem {revtex-au}
W.-J. Jiang, X.-Y. Li, X.-L. Peng, and B.-Q. Sun.
Imaging high-speed moving targets with a single-pixel detector .
Opt. Express 28, 7889-7897 (2020).

\bibitem {revtex-au}
D.-F. Shi, C.-Y Fan, P.-F. Zhang, H. Shen, J.-H. Zhang, C.-H. Qiao, and Y.-J. Wang.
Two-wavelength ghost imaging through atmospheric turbulence.
Opt. Express 21, 2050-2064 (2013).

\bibitem {revtex-au}
Y.-H. Liu, S.-Y. Liu, and F.-X. Fu.
Optimization of Compressed Sensing Reconstruction Algorithms Based on Convolutional Neural Network.
Comput. Sci. 47, 143-148 (2020).


\end{thebibliography}

\end{document}